\documentclass[letterpaper,11pt]{article}
\usepackage[latin1]{inputenc}
\usepackage{amsmath}
\usepackage{amssymb}
\usepackage[american]{babel}
\usepackage{setspace}

\usepackage{vmargin}            
\usepackage[bookmarks,bookmarksnumbered, colorlinks=true, 
		      pdfauthor={Acmae El Yacoubi}]{hyperref}

\author{Acmae El Yacoubi$^{{1}}$, Sheng Xu$^{{2}}$, Z. Jane Wang$^{{1}}$\vspace{10pt}\\
$^{{1}}$ {\it{\small Mechanical and Aerospace Engineering, Cornell University, Ithaca NY, 14853, United States}}\\
$^{{2}}$ {\it{\small Mathematics, Southern Methodist University, Dallas, TX 75275-0156, United States}}}
\title{{\bfseries Falling Particles in Fluids at Intermediate Reynolds Numbers}}

\onehalfspace
\date{}
\begin{document}
\maketitle
\begin{abstract}
In this video, we present the dynamics of an array  of falling particles at intermediate Reynolds numbers. The film shows the vorticity 
plots of 3, 4, 7, 16 falling particles at $Re = 200$.  We highlight the effect 
of parity on the falling configuration of the array.  In steady state, an initially uniformly spaced array forms a convex shape when $n=3$,  i.e the middle particle leads, 
but forms a concave shape when $n = 4$. For larger odd numbers of particles, the final state consists of a mixture of concave and convex shapes. 
For larger even numbers of particles, the steady state remains a concave shape.  Below a threshold of initial particle spacing, particles cluster in groups of 2 to 3.

\end{abstract}
\end{document}